\documentclass[sigconf,10pt, screen, nonacm]{acmart}
\setcopyright{none}

\renewcommand\footnotetextcopyrightpermission[1]{}
\usepackage{algorithm}
\usepackage{algpseudocode}
\algtext*{EndWhile}
\algtext*{EndIf}
\algtext*{EndFor}
\algtext*{EndProcedure}
\usepackage{amsmath}
\usepackage{array}
\usepackage{caption}
\usepackage{color}
\usepackage{comment}
\usepackage[inline]{enumitem}
\usepackage{epsfig}
\usepackage{etoolbox}
\usepackage{fancybox}
\usepackage{float}
\usepackage{fixltx2e}
\usepackage{graphicx}
\usepackage{hyperref}
\usepackage{listings}
\usepackage{multirow}
\usepackage{multicol}
\usepackage{makecell}
\usepackage{placeins}
\usepackage{rotating}
\usepackage{setspace}
\usepackage[hang, tight]{subfigure}
\usepackage{threeparttable}
\usepackage{times}
\usepackage{url}
\usepackage{verbatim}
\usepackage{wrapfig}
\usepackage{tikz}
\usepackage{listings}

\definecolor{codebrown}{rgb}{0.8,0.44,0.2}
\definecolor{codegray}{rgb}{0.5,0.5,0.5}
\definecolor{codepurple}{rgb}{0.58,0,0.82}
\definecolor{backcolour}{rgb}{0.95,0.95,0.92}

\lstdefinestyle{mystyle}{
    backgroundcolor=\color{backcolour},
    commentstyle=\color{codebrown},
    keywordstyle=\color{magenta},
    numberstyle=\tiny\color{codegray},
    stringstyle=\color{codepurple},
    basicstyle=\ttfamily\footnotesize,
    breakatwhitespace=false,
    breaklines=true,
    captionpos=b,
    keepspaces=true,
    numbers=left,
    showspaces=false,
    showstringspaces=false,
    showtabs=false,
    tabsize=2,
    xleftmargin=0.03\textwidth,
    xrightmargin=0\textwidth
}
\newbool{inccomment}
\setbool{inccomment}{true}
\newcommand{\YC}[1]{\ifbool{inccomment}{{\color{magenta}YC\@: #1}}{}}
\newcommand{\XC}[1]{\ifbool{inccomment}{{\color{blue}XC\@: #1}}{}}
\newcommand{\YY}[1]{\ifbool{inccomment}{{\color{orange}#1}}{}}
\newcommand{\FN}[1]{\ifbool{inccomment}{{\color{OliveGreen}#1}}{}}
\newcommand{\GR}[1]{\ifbool{inccomment}{{\color{Tan}#1}}{}}
\newcommand{\LD}{\ifbool{inccomment}{{\color{magenta}\\============================================\\}}}
\newcommand{\RF}{\ifbool{inccomment}{{\color{green}~[R]}}}

\newcommand{\roma}[1]{\uppercase\expandafter{\romannumeral #1\relax}}
\newcommand*\circled[1]{\tikz[baseline=(char.base)]{
            \node[shape=circle,fill,inner sep=0pt] (char) {\textcolor{white}{#1}};}}
\graphicspath{{_fig/}}

\begin{document}

\title[]{GACER: \underline{G}ranularity-\underline{A}ware \underline{C}oncurr\underline{E}ncy \underline{R}egulation for Multi-Tenant Deep Learning
\vspace{2.5mm}
}


\author{Yongbo Yu$^1$, Fuxun Yu$^2$, Mingjia Zhang$^2$, Di Wang$^2$, \\ Tolga Soyata$^1$, Chenchen Liu$^3$, Xiang Chen$^1$}
\affiliation{%
\institution{$^{1}$George~Mason~University, $^2$Microsoft, $^3$University of Maryland}
\country{}}
\email{{yyu25, xchen26}@gmu.edu}

\renewcommand{\shortauthors}{}
\begin{abstract}
\vspace{3mm}

As deep learning continues to advance and is applied to increasingly complex scenarios, the demand for concurrent deployment of multiple neural network models has arisen.
	This demand, commonly referred to as multi-tenant computing, is becoming more and more important.
	However, even the most mature GPU-based computing systems struggle to adequately address the significant heterogeneity and complexity among concurrent models in terms of resource allocation and runtime scheduling.
	And this usually results in considerable resource utilization and throughput issues.
	To tackle these issues, this work proposes a set of optimization techniques that advance the granularity of computing management from both the spatial and temporal perspectives, specifically tailored to heterogeneous model compositions for deep learning inference and training.
	These techniques are further integrated as \textit{GACER} --- an automated optimization framework that provides high-utilization, high-throughput, and low-latency multi-tenant computing support.
	And our experiments demonstrate that \textit{GACER} significantly improves the overall resource utilization and consistently achieves outstanding speedups compared to native GPU computing frameworks and existing state-of-the-art optimization works.

\end{abstract}


\maketitle

\section{Introduction}
\label{sec:intro}
\vspace{3mm}

The explosive success of deep learning techniques in various cognitive tasks, such as image classification and speech recognition, has made neural network models the hot spot of computing systems research.
	One of the primary drivers behind this trend is the support provided by GPUs, which offer excellent computing capacity and parallelism capability~\cite{owens2008gpu,keckler2011gpus}.
	Despite of the active emergence of various ASIC- and FPGA-based accelerators or systems, GPUs remain the most widely adopted platform in practice, accounting for >85\% market share in both cloud and edge applications~\cite{gpu_report}.

While, for a long time in the past, due to substantial parameters of neural network models, many research works tended to adopt a generic setting: one GPU instance could only host a single model.
	And even many recent outstanding efforts have not stepped out of this rut (e.g., MetaFlow~\cite{jia2019optimizing}, IOS~\cite{ding2021ios}).
However, cutting-edge developments have gradually revolutionized this setting:
	With the miniaturization of neural networks and the massification of GPUs, it is now possible for a single GPU to host multiple models simultaneously~\cite{lin2019towards,howard2017mobilenets}.
	Furthermore, the need for concurrent model processing also scales up with multi-task or multi-modality intelligence integration, such as in autonomous driving~\cite{liu2019edge,mystakidis2022metaverse}.
	Along with this trend, GPU manufacturers like NVIDIA actively promote related software and hardware techniques~\cite{MPS, Multi-Stream, MIG}.
	Thus, ``\textbf{multi-tenant}'' deep learning computing becomes more and more prominent, especially for GPU-based systems.

However, multi-tenant deep learning presents a greater challenge than conventional single-tenant deployment regarding computing management.
	This is due to the increased heterogeneity and complexity of the concurrency of multiple neural network models, which can vary in operator composition, structural design, and model scale~\cite{yu2022powering,choi2021lazy}.
	Unfortunately, these issues are not well addressed in current GPU computing, even with emerging supporting techniques from manufacturers:
	(1) When it comes to resource allocation, current techniques for issuing concurrent models into the GPU either result in using fixed hardware resource budgets or competing models for resources in a greedy manner.
	These approaches at the model level can lead to wastage of hardware capabilities or resource contention and corresponding overhead~\cite{koller2019weakly,kwon2020nimble,dhakal2020gslice}.
	(2) When it comes to runtime scheduling, though many recent works have dived into the operator level, most of them either cannot handle multi-tenant scenarios or have overlooked the multi-tenant coordination overhead, leading to ineffective and unscalable runtime management~\cite{ding2021ios,yu2021automated}.

%

Given these observations, the primary motivation for multi-tenant deep learning optimization is to develop a fine-grained and feasible coordination computing framework to resolve resource contention and enhance complementary resource utilization across different model tenants with minimal regulation overhead.
	Therefore, several optimization expectations for multi-tenant deep learning can be derived:
	(1) From a spatial perspective, the resource allocation scheme requires unprecedented operator-level granularity to adapt to the dynamics of intra-model operators and therefore complement intra-model resource requirements~\cite{xu2016warped,yu2021automated}.
	(2) From a temporal perspective, the granularity of multi-tenant runtime scheduling should not only deepen to the operator level, but also improve the manageability to regulate the runtime overhead and balance the overall performance given different deployment complexities~\cite{yu2021automated,choi2021lazy}.
	These two optimization expectations from both the spatial and temporal perspectives point to the same optimization focus of this work:
	\textit{To advance the current multi-tenant concurrency regulation granularity and co-optimize the spatial and temporal domains for optimal computing performance}.

\vspace{2mm}
Centering on this research focus, our work extensively examines the entire GPU-based system stacks and conducts a comprehensive analysis on current multi-tenant computing issues with state-of-the-art techniques.
	Subsequently, we propose a set of optimization techniques that advance the granularity of computing management in both the spatial and temporal management domains, significantly improving runtime performance for both deep learning inference and training.
	The contributions of this work are as follows:

\vspace{-1mm}
\begin{itemize}[noitemsep,leftmargin=*]
		\vspace{0.9mm}
	\item We reveal computing issues in multi-tenant deep learning and formulate corresponding processes for GPU-based systems and deep neural network (DNN) tenants.
		And this allows us to further identify multi-tenant optimization objectives and approaches.
		\vspace{0.8mm}
	\item In the spatial domain, we propose a set of novel operator resizing and decomposition methods combined with emerging GPU techniques to enhance the granularity and flexibility of multi-tenant resource allocation.
		It fully resolves the model contention and exploits resource utilization.
		\vspace{0.8mm}
    \item In the temporal domain, we improve multi-tenant scheduling at the operator level by optimizing related operator issuing and CPU-GPU synchronization mechanisms.
		Additionally, we identify specific scheduling overheads and regulate corresponding optimization trade-offs to improve the runtime performance comprehensively.
		\vspace{0.8mm}
	\item The proposed techniques are integrated into an automated optimization framework --- \textit{GACER}, which leverages a low-cost search method to identify the particular spatial and temporal deployment configuration for both offline and online multi-tenant deep learning scenarios.
		\vspace{0.7mm}
\end{itemize}

	\textit{GACER} could consistently provide high-utilization and high-throughput computing support to multi-tenant deep learning on GPUs.
	Compared to conventional computing framework without specific multi-tenant support (e.g., TVM), \textit{GACER} could consistently achieve almost $\sim$70\% speeds up.
	And comparing to state-of-the-art multi-tenant optimization works, it could also achieve $\sim$30\% acceleration with $\sim$40\% resource utilization enhancement, and demonstrate outstanding capabilities for even  more complex deployment scenarios.

\begin{figure}[!t]
	\centering
	\captionsetup{justification=centering}
	\hspace{5mm}\includegraphics[width=3.3in]{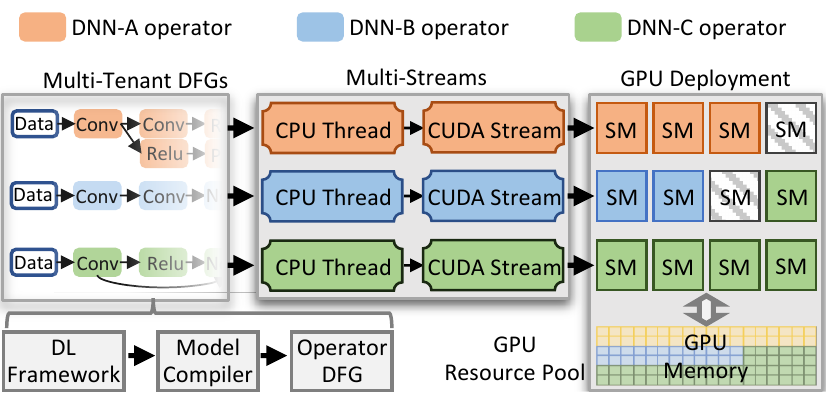}
	\vspace{-0.3mm}
	\caption{Multi-Tenant Deep Learning with GPUs\vspace{-2mm}}
	\label{fig1-deployment}
\end{figure}

\section{Preliminary}
\label{sec:prelim}

\subsection{Multi-Tenant Deep Learning}
	This paper focuses on a generic multi-tenant deep learning deployment setting on GPUs~\cite{choi2021multi}, as shown in Fig.~\ref{fig1-deployment}
	\footnote{Note that the techniques proposed in this paper are applicable to both the training and inference phases of multi-tenant deep learning.
	Hence, we will not make special distinctions in the following descriptions.}:

\renewcommand{\arraystretch}{1.5}
\begin{table*}[!t]
\setlength{\tabcolsep}{1.3mm}
	\centering
	\footnotesize
    \vspace{-4mm}
	\caption{Granularity and Performance of Recent Works on Multi-Tenant DL Computing Optimization}\label{tab:recent method}
	\vspace{-3mm}
	\begin{tabular}{c|cllc}
		\hline
		& \makecell[c]{\textbf{Techniques}} & \makecell[c]{\textbf{Approaches}} & \makecell[c]{\textbf{Granularity}} & \textbf{Remaining Issues} \\
		\hline
		\multirow{4}{*}{\rotatebox{90}{\centering \textbf{\textit{Spatial}}\hspace{6mm}}}
		& \makecell[c]{Multi-Instance GPU\\\makecell[c]{(MIG)~\cite{MIG}}} & \makecell[l]{$\bullet$~Physical GPU Instance Partition\\$\bullet$~Instance-based Tenant Allocation} & \makecell[c]{$\bullet$~Model Level} & \makecell[l]{$\bullet$~Fixed Instance Budget\\$\bullet$~Reconfiguration Overhead} \\
		\cline{2-5}
		& \makecell[c]{Multi-Process Service\\\makecell[c]{\hspace{-2mm}(MPS)~\cite{MPS}}} & \makecell[l]{$\bullet$~Virtualized GPU Resource Partition\\$\bullet$~Dynamic Process Resource Budget} & \makecell[c]{$\bullet$~Model Level} & \makecell[l]{$\bullet$~Resource Contention\\$\bullet$~Context Switch Overhead} \\
		\cline{2-5}
		& \makecell[c]{Gslice\cite{dhakal2020gslice},~Salus\cite{yu2020salus}} & \makecell[l]{$\bullet$~Multi-Tenant with MPS\\$\bullet$~Batching Configuration\\$\bullet$~Overhead Optimization} & \makecell[c]{$\bullet$~Model Level} & \makecell[l]{$\bullet$~Coarse-Grained Optimization} \\
		\cline{2-5}
		\hline
		\multirow{3}{*}{\rotatebox{90}{\centering \textbf{\textit{Temporal}}\hspace{4mm}}}
		& \makecell[c]{Time-Sliced Sharing\\(EdgeBatch~\cite{zhang2019edgebatch}, Gpipe~\cite{,huang2019gpipe})} & \makecell[l]{$\bullet$~Runtime Singe-Tenant Switch} &\makecell[c]{$\bullet$~Sub-Model/\\Model Level} & \makecell[l]{$\bullet$~Tenant Switch Overhead} \\
		\cline{2-5}
		& \makecell[c]{Multi-Stream\\\makecell[c]{(MS)~\cite{Multi-Stream}}} & \makecell[l]{$\bullet$~Post-Compilation Stream Resource Allocation\\$\bullet$~CPU-GPU Synchronization based Scheduling} & \makecell[c]{$\bullet$~Model Level} & \makecell[l]{$\bullet$~Resource Contention\\$\bullet$~Greedy Runtime Management} \\
		\cline{2-5}
		& \makecell[c]{Operator Scheduling\\(IOS\cite{ding2021ios},~AutoMT\cite{yu2021automated})} & \makecell[l]{$\bullet$~DFG-based Scheduling with MS\\ $\bullet$~Resource Contention Optimization}  & \makecell[c]{$\bullet$~Operator Level} & \makecell[l]{$\bullet$~Unregulated Resource Allocation\\$\bullet$~Runtime Regulation Overhead} \\
		\hline
	\end{tabular}
	\vspace{-2mm}
\end{table*}

\vspace{1mm}
\textbf{Multi-Tenant Heterogeneity:}
When multiple heterogeneous DNN models are deployed on a single GPU, and each is compiled into a data flow graph (DFG) that defines the computing sequence of a series of different operators by layers (e.g., Conv, ReLu, etc.).
	And each operator has a particular computational pattern and resource requirements, including computing resources measured by the occupancy of streaming multi-processors (SM), and memory resources measured by memory bandwidth utilization.
	Despite the particular operator design, the resource consumption of each operator is mainly determined by the batched job sizes.

\vspace{1mm}
\textbf{Multi-Tenant Deployment:}
The DFG compilation is generally performed on the CPU side, and the DFG of each DNN tenant is wrapped into a processing thread for later GPU deployment.
	It's worth noting that the simultaneous multi-thread wrapping process has much less overhead than initializing operators one at a time.
	On the GPU side, each thread is assigned to a specific GPU computing stream \cite{Multi-Stream} with a dedicated resource portion from the GPU's SM pool.
	As shown in Fig.~\ref{fig1-deployment}, such a concurrent multi-tenant deployment significantly differs from traditional time-sliced deployment, where only one model runs at a time.

\vspace{1mm}
\textbf{Resource Sharing and Contention:}
As multiple models are deployed concurrently and share the same GPU pool, resource sharing has become a major research focus.
	And various resource allocation and management schemes have been proposed to improve the complementary resource utilization across different model tenants, as will be reviewed in the next section.
	However, such resource-sharing mechanisms also lead to potential resource contention~\cite{yu2021automated} among tenants when they compete for limited GPU resources, which can result in certain performance issues such as latency increment and throughput reduction.
	And when we further take into account the heterogeneity between DNN tenants, multi-tenant computing support becomes even more challenging.

\subsection{Multi-Tenant Computing Support}
Table~1 reviews state of the art for multi-tenant computing.

\vspace{1mm}
\textbf{Emerging GPU Techniques:}
New GPU architecture features proposed by NVIDIA, such MIG~\cite{MIG} and MPS~\cite{MPS}, have enabled GPUs to divide the SM pool into multiple concurrent portions for multi-tenant deployment.
	While these methods provide GPUs with spatial resource management capabilities, they often suffer from reconfiguration overhead and a lack of certain runtime flexibility.

	Unlike MIG and MPS, which partition resources first and then deploy tenants, MS enables tenant-oriented dynamic resource allocation at runtime, facilitating post-compilation resource budget adjustment for multi-tenant models.
	Furthermore, MS supports frequent CPU-GPU synchronizations, which enable DFG reordering along stream issuing, thus providing the scheduling management capability from a temporal perspective.
	However, coordinating such multi-tenant GPU support is often overwhelming.
	Facing the multi-tenant complexity, MS usually simply adopted a greedy manner of runtime management, resulting in considerable resource contention and inappropriate scheduling cases~\cite{yu2021automated}.

\begin{figure}[!b]
	\centering
	\captionsetup{justification=centering}
	\vspace{-5mm}
	\hspace{5mm}\includegraphics[width=3.3in]{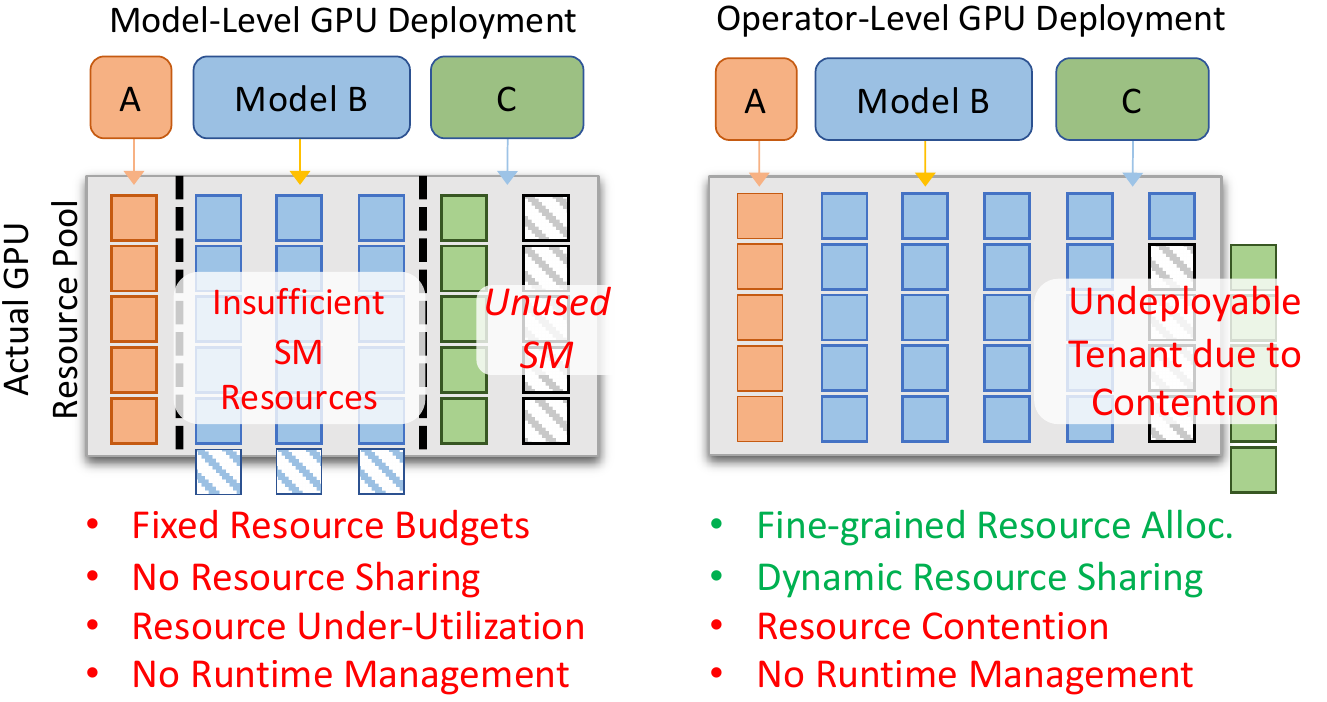}
	\vspace{-2.2mm}
	\caption{Comparison of Different Granularity\vspace{0.5mm}}
	\label{granularity_compare}
\end{figure}

\vspace{1mm}
\textbf{State-of-the-Art Optimization Works:}
Based on these emerging techniques, many optimization works have been proposed.
	In the spatial domain, some studies have combined model-level workload/resource adaption with MIG or MPS to implement software-hardware co-optimization for enhancing multi-tenant resource sharing~\cite{dhakal2020gslice,yu2020salus}.
	In the temporal domain, very recent works have expanded the computing granularity of MS into the operator level.
	However, these works are still not comprehensive.
	Most of them either cannot handle multi-tenant scenarios or have overlooked the multi-tenant coordination overhead, leading to ineffective and unscalable runtime management~\cite{ding2021ios,yu2021automated}.

\begin{figure*}[!t]
	\centering
	\vspace{-4mm}
	\includegraphics[width=6.8in]{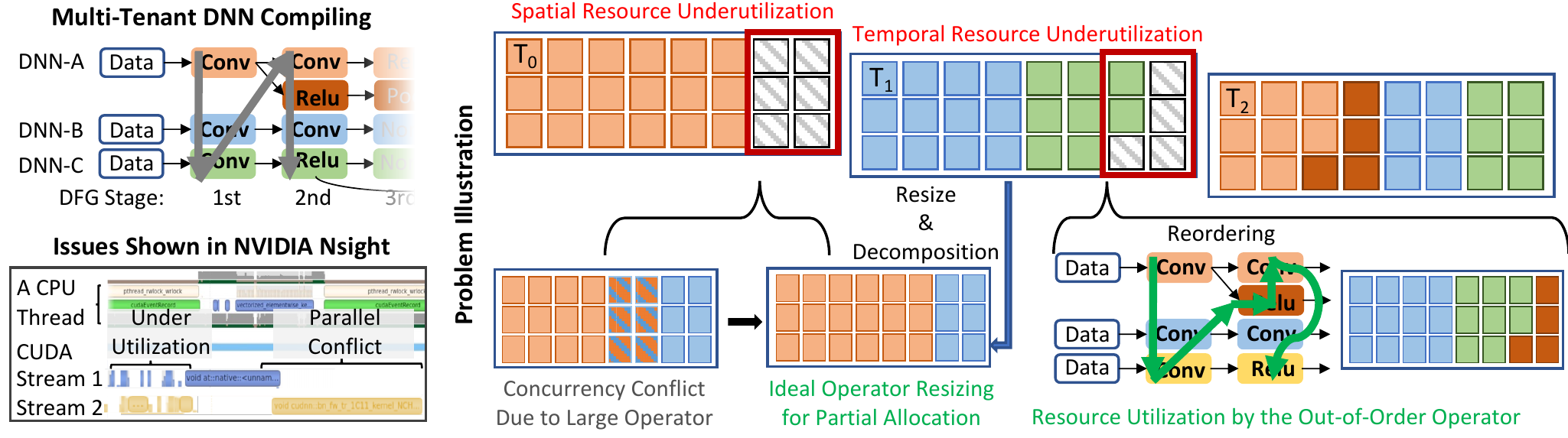}
	\vspace{-2mm}
	\caption{GPU Resource Utilization Analysis from Spatial and Temporal Perspectives}\label{fig:residue}
    \vspace{-2mm}
\end{figure*}

\subsection{Multi-Tenant Computing Granularity}
In addition to reviewing the basic mechanisms, pros and cons of different techniques, Table 1 also provides a comparison of the optimization granularity.
	While these works have enabled multi-tenant deep learning on GPUs, most are still limited to a coarse granularity (esp. at the model level) and focus on a singular optimization domain (either spatial or temporal).
%

Fig~\ref{granularity_compare} presents a brief comparison to demonstrate the impact of optimization granularity:
	(1) When multi-tenant DNNs are deployed with only model-level granularity, SMs are partitioned based on an average model resource requirement, as in the case of MPS and MIG.
	However, when considering individual operators by layer during runtime, the specific resource requirements may vary, resulting in resource underutilization with smaller operators or insufficient resources with larger ones, requiring additional GPU cycles.
	(2) As shown in Fig~\ref{granularity_compare}, a more fine-grained resource allocation at the operator level could overcome the fixed resource budgets and accommodate the varying resource requirements at individual model layers.
	However, due to the significant heterogeneity of multi-tenant models, significant resource contention can still occur.
	Although this could be addressed by the algorithm and compiling optimization (e.g., TVM~\cite{chen2018tvm}) or runtime scheduling (e.g., AutoMT~\cite{yu2021automated}), additional efforts are still required to refine the dynamic resource allocations and related overhead.

\section{Design Analysis and Motivation}
Therefore, this work focuses on advancing the manageability and efficiency of multi-tenant deep learning and at the fine-grained operator-level and expanding the spatio-temporal optimization space for comprehensive performance escalation.
	This section presents a detailed analysis of GPU resource utilization during multi-tenant DNN deployment and discusses the underlying design methodology of this work.

\subsection{Persistent Resource Utilization Issues}
Fig.~\ref{fig:residue} highlights common multi-tenant computing issues that arise when deploying heterogeneous DNN models on a GPU with MS techniques and the ideal operator-level resource allocation.
	In this process, each operator is compiled with a defined resource budget and subsequently issued in a greedy order, as illustrated by the gray arrows in the DFG compiling figure.
	Such a subsequently issuing will allocate operators from different model tenants into individual CUDA streams to share the computing resource at each cycle.

Despite the utilization of the best available multi-tenant deployment settings above, resource underutilization issues  still inevitable.
	In most deployment scenarios, even with operator-level resource allocation in place, the total resource requirements from concurrent tenants of a DFG stage will not exactly match the available GPU resource volume.
	As mentioned earlier, when there is a deficiency of any of the required resources (e.g., SM or memory) to deploy a model tenant's operator, the operator is moved to the next cycle.
	To illustrate this persistent issue, the lower-left figure displays an empirical example using the NVIDIA Nsight~\cite{Nsight}.
	As shown there:
		In some GPU cycles, the concurrently deployed operators are unable to fully utilize the SM pool; and in some other cycles, concurrent operator deployment cannot be established for parallel processing, resulting in even worse resource underutilization.

\vspace{-4mm}
\subsection{Design Motivation and Challenges}
Here we refer to the unused portion of the GPU resources as ``residue'', and a GPU SM pool-oriented example is shown in the right part of Fig.~\ref{fig:residue}, which is further analyzed from both spatial and temporal perspectives.

\vspace{1mm}
\textbf{Spatial Residue Optimization by Operator Resizing:}
As aforementioned, a spatial optimization approach could be adjusting the deployed model tenant workload to accommodate resource availability with less contention~\cite{yu2020salus}.
	And it will be even more effective when pushing the granularity from the model level to the operator level for operator resizing.

	However, certain challenges arise:
	(1) Batching operation is always applied to an entire model in the current frameworks (e.g., PyTorch~\cite{pytorch}).
	Since the total workload of individual operators is invariant, operator resizing can only follow the path of decomposing the workload into relatively small operator units.
	However, conventional operator decomposition remains at the model compilation stage and cannot satisfy the computational dynamics of multi-tenant scenarios, requiring a necessary framework modification.
	(2) Meanwhile, in multi-tenant scenarios, the optimization complexity also scales up to determine the appropriate operator for resizing, taking into account the runtime overhead.
	Thus, a comprehensive multi-tenant-specific optimization scheme is also required.

\vspace{1.5mm}
\textbf{Temporal Residue Optimization by Operator Reordering/Scheduling:}
As shown in Fig.~\ref{fig:residue}, the DFG reordering indicated by the green arrows could also exploit the underutilized GPU residual.
	In addition to conventional scheduling issues, such as layer dependency, new challenges are posed:
	(1) Multi-tenant scheduling involves multiple operators across various DFG settings, resulting in unprecedented dynamic complexity.
	Although the CPU-GPU synchronization provided by MS could achieve model reordering, specific library-level optimization is required to decompose the model DFG to the operator level for finer granularity.
	(2) Not coincidentally, significant DFG reconfiguration delay would be introduced, requiring an effective overhead regulation scheme.

\vspace{1.5mm}
\textbf{Spatial and Temporal Co-optimization}
It is important to note that in a residue case, both spatial and temporal optimization methods can be adopted.
	Therefore, determining the trade-off between these methods is critical in selecting the appropriate one.
	Furthermore, a series of optimization procedures through multi-tenant computation process are interdependent.
	In other words, the solution of one residue may affect the subsequent ones, making the previously determined locally optimal solution inefficient.
	Hence, this complexity renders spatial and temporal co-optimization as the most crucial aspect of this work.

\section{Granularity-Aware Multi-Tenant Regulation}
\label{sec:4}

\subsection{Problem Formulation}
\begin{figure}[!b]
	\centering
	\vspace{-4mm}
	\includegraphics[width=3.3in]{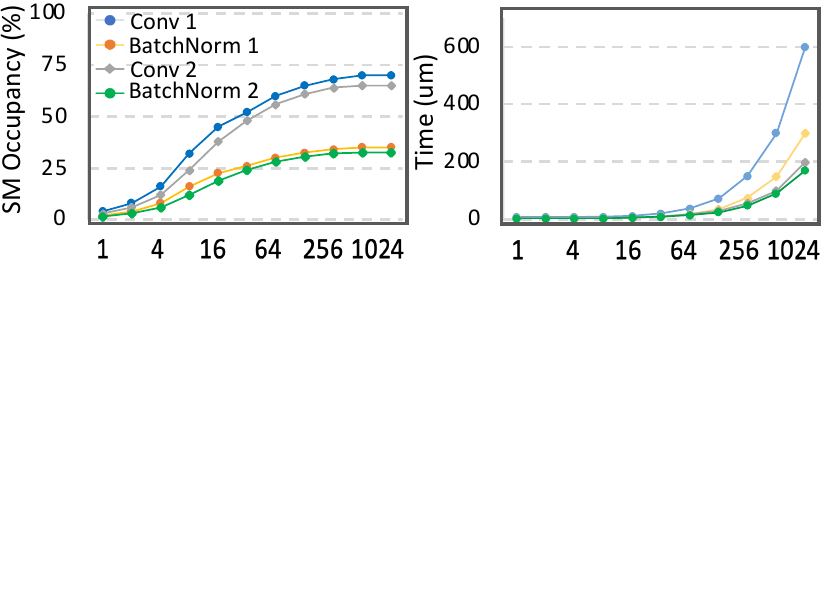}
	\vspace{-36mm}
	\caption{Resource Utilization and Time Profiling}
	\vspace{-1mm}
	\label{fig:resource_batch}
\end{figure}

\hspace{\parindent}
\textbf{Resource and Tenant Formulation:} 
 We abstract the total GPU SM resource pool $S_{GPU} = 100\%$.
    And we map the operator workload $W(O^B)$ to the SM occupancy, where $O$ is the operator, $B$ is the batch size of $O$
    We analyze DNN operators with different batch settings and formulate a lookup table for convenience, and give some examples of convolution and batchnorm operators in Fig.~\ref{fig:resource_batch}.
    The execution time $T$ required by operators $O$ also varies, and some operators need to span multiple time cycles for processing, so we also formulate the operator execution time.

\vspace{1mm}
\textbf{Multi-Tenant Runtime Deployment:}
Multi-tenant DNN consists of several parallel tenants, and all tenants consist of $N$ models: $M_1, M_2,..., M_n$.
    And each model $M$ can be represented by a DFG with a series of operators $O$.
    So we have the model operator list $M_n = [O_{n,1},O_{n,2},...,O_{n,i}]$.
    
The aforementioned formulation primarily illustrates the spatial mapping of resources and tenants, wherein a resource pool is utilized. 
    This temporal deployment during runtime may be expressed for each cycle as follows:
\begin{equation}
    \begin{split}
	\small
	& S_{T_0} : [O_{1,1}], ~~~ S_{T_1} : [O_{2,1}, O_{3,1}], 
    \\& S_{T_2} : [O_{1,2}, O_{1,3},O_{2,2}, O_{3,2}], ..., S_{T_t} : [O_{n,i},...]
    \\& where~S_{T_0}, S_{T_1},..., S_{T_t}<= S_{GPU}.
	\label{eq:sequence}
    \end{split}
\end{equation}
    In each time cycle, the aggregate SM occupancy $S_T$ of operators must be maintained below the $S_{GPU}$. 
    This necessitates the deployment of certain operators across multiple time cycles, as exemplified by the notation $S_{T_0 \rightarrow T_3}:[O_{1,1}]$.
    This time cycle deployment essentially reflects the operator execution sequence $S_{T_0} \rightarrow S_{T_t}$.

\vspace{1mm}
\textbf{Optimization Objective:}
Upon the GPU resource and tenant formulation above, the residual resource in time cycle $S_T$ mentioned in Section 3.3 could be formulated as follow:
\begin{equation}
\small
R_{S_T}= S_{GPU}-S_T = S_{GPU}-\sum_{O~in~S_T}W(O_{n,i}^{B_{n,i}}).
\label{eq:R_S}
\end{equation}
And we subsequently sum the $R_{S_T}$ across all cycles to compute the total residue $R$.
\begin{equation}
	\small
    R= \sum_{S_{T_0} \rightarrow S_{T_t}}(S_{GPU}-\sum_{O~in~S_T}W(O_{n,i}^{B_{n,i})}).
	\label{eq:R_batch}
\end{equation}
$R$ has two variables: operator batch size $B_{n,i}$ and operator execution sequence $S_{T_0}\rightarrow S_{T_t}$, which are both in operator-level granularity.
Therefore, we could optimize two sub-objective, finding the appropriate $B$ and $S_{T_0} \rightarrow S_{T_t}$ to minimize the $R$.
These could be translated into two sub-objective, finding the appropriate $B$ and $S_T$ to minimize the $R$.
\begin{equation}
	\small
    \tau~(B_{n,i},~~S_{T_0} \rightarrow S_{T_t}) \rightarrow Min~R,
	\label{eq:min_R_batch}
\end{equation}
where $\tau$ is an approximation approach.

Based on the two variables of Eq.~\ref{eq:min_R_batch}, we need to extend the spatial and temporal granularity of the multi-tenant DNN to the operator level for solving $R$.
In the following context, we introduce the spatial and temporal regulation methods to adjust $B_{n,i}$ and $S_{T_0} \rightarrow S_{T_t}$ to fine granularity and present an optimal search strategy for $\tau$ implementation.

\subsection{Spatial Granularity Regulation}

\hspace{\parindent}
\textbf{Regulation with Operator Resizing:}
Instead of deploying all workloads of a certain operator to the GPU in the same time cycle, they can be resized into multiple smaller copies by decomposing the operator through batch size direction. 
\begin{equation}
\begin{split}
	\small
	O_{n, i}^B\xrightarrow{\hspace{2mm}\text{chunk}\hspace{1mm}} & O_{n, i}^{B^1}, O_{n, i}^{B^2},...,O_{n, i}^{B^j}, ~~\text{where}\sum_{1~\text{to}~j} B^j = B.
	\label{eq:resizing}
\end{split}
\end{equation}
The decomposed operators can be deployed to different time cycles, thus, we could only deploy part of the workload in a single time cycle.
    Therefore, the first objective could be translated into finding the batch size $list_{B_{n,i}}=[B^1, B^2,...B^j]$ of each operator $O_{n, i}$.
    By controlling the number $j$, we can control the spatial granularity of each operator deployment. 

\vspace{1mm}
\textbf{Regulation with Operator Decomposition}
The current DNN model API structure employs a single batch size for all layers, resulting in a model-level granularity that fails to consider varying workloads in each layer. 
To address this issue, we used PyTorch's "torch.chunk()" and "torch.cat()" functions to rewrite the API model definition. 
By decomposing heavy workload operators into smaller ones and concatenating the resulting micro-batches, we can effectively control runtime resource consumption without sacrificing model accuracy. 
This approach increases spatial granularity and allows for finer-grained parallelism. Resizing the batch size in this manner enables us to adjust the workload at a more granular level, leading to optimized performance.

\vspace{1mm}
\textbf{Resizing Regulation Overhead Analysis:}
The resizing regulation needs to introduce additional decomposing and concatenation operations which also bring additional overhead.
    This makes the decomposed operator also have the trade-off between residue reduction and additional overhead.
    Therefore, the decomposed operator should be near the large residue in order to bring enough gain.
    However, not all operators in multi-tenant DL computing require decomposing along the batch direction. 
    We design a mask list whose length is equal to the total number of operators in all DFGs, and each element in the mask is corresponding to an operator.
    If the $mask(O_{n, i})$ is 0, that indicates $O_{n,i}$ is no need to decompose, otherwise, a $list_{B_{n, i}}$ is generated for $O_{n, i}$ to represent how it was decomposed.
    During the optimization process, we will gradually find which operators need to be decomposed.

\vspace{1mm}
\textbf{Overall Spatial Regulation:}
Based on the above overhead analysis, we calculate the biggest residue $Max(R_{S_T})$ using Eq.~\ref{eq:R_S} and decompose the operator with the largest size following this time cycle.
After that, we decompose a batch that matches the residue size and update the mask list and $list_B$ and update the decomposed operators to the DFG.
And we check the dependence to make sure the decomposed operator could be deployed in this residue.
However, due to the unequal length of some segments in the same cluster, operators in the tail of the longest segment have to execute individually and inevitably generate a larger residue.
These residues do not need to be optimized, so we skip them. 
 
Overall, the first sub-objective of spatial granularity regulation could be translated into finding the decomposition mask matrix and corresponding decomposition strategy $list_B$.

\begin{figure}[!t]
	\centering
	\captionsetup{justification=centering}
	\includegraphics[width=3.3in]{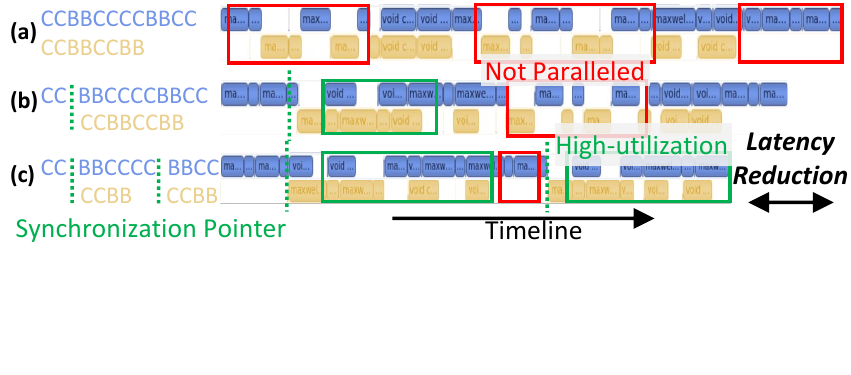}
	\vspace{-14mm}
	\caption{Reordering with Synchronization Pointers\vspace{-3mm}}
	\vspace{-2mm}
	\label{fig:case}
\end{figure}

\subsection{Temporal Granularity Regulation}

In the temporal domain, we break through multiple DFGs into fine-grained operator segments, and therefore operator scheduling across multiple different models is implemented.

\vspace{1mm}
\textbf{Operator Level DFG Reordering:}
In order to reorder the operator execution sequence $S_{T_0}\rightarrow S_{T_t}$, we insert some pointers in DFG, and divide each DFG into several segments with different granularity (e.g. operator, stage, and the whole model), so that we could implement inter-model concurrent scheduling with different granularity.
And then, we control which segments can be deployed simultaneously.
By changing which segment the operator is located in, the order of execution of the operator is relatively changed.
As shown in Fig~\ref{fig:case}, we use some demo models to show what is the pointer and how operator reordering could improve the parallelism performance, and each model is deployed in each stream. 
    For simplicity, each sequence only consists of one type of the two operators $C$ (e.g., Conv) and $B$ (e.g., BatchNorm).
    As shown in Fig~\ref{fig:case} (b),  by inserting the pointer to reorder the execution sequence, the first operator $C$ in stream 2 is able to perform concurrency with $B$ in stream 1. 

We formulate this process as follows,
and taking stream 1 in Fig.~\ref{fig:case} (c) as an example, the model $M_1$ has 12 operators and is divided into three segments using two pointers.
And then, the same index segments from different models are divided into the same cluster which can be deployed simultaneously, such as the cluster in Equation~\ref{eq:issue_grid_result} $S_{T_3 \rightarrow T_8}$.
The operators in the same cluster can occupy multiple time cycles (assuming that all operators occupy only one cycle in this case).
\begin{equation}
\begin{split}
\small
& S_{T_0 \rightarrow T_2}: [O_{1,1},O_{1,2}], [None], \\
& S_{T_3 \rightarrow T_8}: [O_{1,3},...,O_{1,8}],[O_{2,1},...,O_{2,4}], \\
& S_{T_9 \rightarrow T_{12}}: [O_{1,9},...,O_{1,12}],[O_{2,5},...,O_{2,8}].
\label{eq:issue_grid_result}
\end{split}
\end{equation}

Since the total time cycle of each DFG is not the same in the multi-tenant DL scenarios. 
Therefore, we do not need and cannot ensure that each segment is of equal length.
We object to minimizing the overall time cycle occupied, which means that a smaller residue is generated.

Therefore, the operator execution sequence $S_{T_0}\rightarrow S_{T_t}$ regulation could be translated to find the best segments and the optimal scheduling strategy. 
We use a synchronization pointer to detach each model’s full sequence into several operator segments and establish a pointer matrix $Matrix_P=[P_1,P_2,...P_n]$.
Each model $M$ has a list $P$, and each $P$ annotates the appropriate positions where we detach the DFG.
\begin{equation}
\begin{split}
\small
M_1:[O_{1,1},O_{1,2},...,O_{1,12}] + P_1:(2,8) = Seg(M_1).
	\label{eq:model_issue_grid}
\end{split}
\end{equation}
Each number in $P$ represents the position at which the pointer is inserted. 
Thus, the second sub-objective could be translated into finding the pointer matrix $Matrix_P$ for each DFG.
Each $P$ has the same number of pointers.

\begin{figure}[!t]
	\centering
	\includegraphics[width=3.3in]{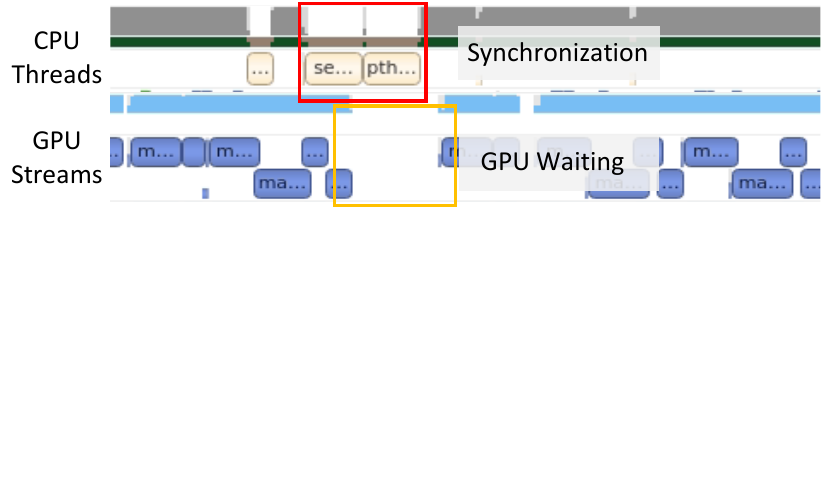}
	\vspace{-28mm}
	\caption{CPU-GPU Synchronization Overhead}
	\vspace{-3mm}
	\label{fig:sync_overhead}
\end{figure}

\vspace{1mm}
\textbf{Runtime Synchronization Modification at Library Level:}
The synchronization pointer is a GPU synchronization mechanism, that ensures that operators continue to be deployed only after previously issued operators have completed their computation.
Through these pointers, we could control which operators could be deployed simultaneously.
However, the current DNN model structure makes the scheduler only take the whole model as a scheduling unit, which causes course-grained temporal granularity.
This prevents us from playing the role of optimizing temporal granularity with synchronous pointers.
Therefore, we also redesign the DNN API at the library level.
Based on the PyTorch framework, we use the "model.named\_modules()" method to get all the model layer objects and use "nn.Sequential" to refine all the objects into a list of DFG objects. 
Based on the above two operations, we are able to implement operator reordering and thus fine-grained scheduling of operators.

\vspace{1mm}
\textbf{Temporal Regulation Overhead Analysis:}
However, the reordering method uses synchronization operation which is the runtime control, so this inevitably introduces scheduling overhead as shown in Fig.~\ref{fig:sync_overhead}.
    The GPU waits until the CPU finishes synchronizing the pointer and then sends a new operator to GPU. 
    This situation can lead to a lot of GPU wait time, resulting in additional residual resources.
    Adding a large number of pointers leads to frequent GPU waits, which can seriously slow down the overall efficiency of the GPU.
    So for different scenarios, we also need to trade off the gains and overheads that come with temporal granularity.

\vspace{1mm}
\textbf{Overall Temporal Regulation:}
Based on the above overhead analysis, we readjust the computation of the residue so that the solution process can achieve granularity awareness.
    This waiting time is equivalent to introducing an additional residue of $S_{GPU} *$ GPU synchronization wait time $T_{SW}$.
    In the same computer system, this overhead is relatively stable and we can obtain roughly accurate values by profiling.
    Therefore, we can add the following term to Eq.~\ref{eq:min_R_batch} to adjust the objective function to be aware of the overhead caused by fine-grained scheduling.
\begin{equation}
	\small
    R= \sum_{S_{T_0} \rightarrow S_{T_t}}(S_{GPU}-\sum_{O~in~S_T}W(O_{n,i}^{B_{n,i}})) + |P_n|*S_{GPU}*T_{SW},
	\label{eq:min_R_batch_overhead}
\end{equation} 
where $|P_n|$ is the number of pointers.
When $|P_n|$ is larger, the antecedent term has a greater potential to obtain a smaller residue but will make the posterior term larger.
This makes it profitable to fill only larger residues, which tend to occur in the first few time cycles of larger operators.


\subsection{Granularity-Aware Joint Optimization}

\hspace{\parindent}
\textbf{Search Framework for Joint Optimization:}
Based on the formulation and regulation design, we propose our granularity-aware framework to minimize $R$ by finding the optimal strategy from the search space of the mask, $list_B$, and $Matrix_P$.

We have three claims in our framework.
(1) We do joint optimization of spatial and temporal granularity.
    And the adjustable factors in search space are coupled for the effect of residue, so finding the globally optimal solution is NP-hard.
    Therefore, we greedily alternate between spatial and temporal regulation until we reach the optimal concurrency strategy.
(2) When doing joint optimization, we do not only consider the SM resource pool, but we can also extend this approach to other resources, such as GPU memory bandwidth.
(3) Our framework also needs to be aware that fine-grained granularity optimization brings gain while also introducing overhead. 
    we need to implement different granularity for different scenarios.

We propose a search approach that could consider all three claims and further propose a framework based on this search method as shown in Algorithm 1.
We first initialize the elements in $mask(O)$ and $Matrix_P$ to 0 and calculate the residue using Eq.~\ref{eq:min_R_batch_overhead}.
For spatial granularity, we use the spatial regulation method in section 4.2.
    For temporal regulation, we implement a coordinate descent search algorithm, which takes the pointer number in $Matrix_P$ as different coordinates.
    Then the optimal pointer number for each list $P$ is searched alternately, during which other $P$ lists are kept as the previous optimal one. 
    The Spatial and Temporal search method is executed alternately. 
    After a certain round of temporal execution, we switch to the spatial regulation search method in section 4.2 and update the DFG.
    And the decomposed operators are inserted between the pointers, without affecting the scheme of the existing $Matrix_P$.
\begin{algorithm}
		\caption{Granularity-Aware Search}  
		\begin{algorithmic}[1] 
			\Require $N$ DFGs, the initialized mask list, $list_B$ and $Matrix_P$, the rounds of search $X$.
			\Ensure The optimal mask list, $list_B$, and $Matrix_P$
			\State Initialize a dictionary $D\{R:Matrix_P\}$ to record residue $R$ and the corresponding $P$.
            \For{rounds $=x$  to $X$ }                
                \For{model $i=1$  to $N$}
                    \For{the $j$  in $P_j$}
                        \State Calculate $R$ using equation~\ref{eq:min_R_batch_overhead}
                        \State Append ${R:Matrix_P}$ to the records $D$.
                    \EndFor
                    \State Update the $P_j$ of $Matrix_P$ with the smallest $R$.
                \EndFor
            \EndFor
            \State Sort the records $D$ by the $R$. 
            \If{the smallest $R$ in $D_{|P_n|}$ $>$ $R$ in $D_{|P_n|-1}$}
                \State \Return $Matrix_P$ with the smallest $R$ in $D_{|P_n|-1}$.
            \EndIf
            \State Add pointer in $Matrix_P$
            \State Go back to Step 2
		\end{algorithmic}  
\end{algorithm}

Our framework --- \textit{GACER}, is a functional addition to the mainstream framework with good scalability and can be extended to a variety of applications at a low cost.

\vspace{1mm}
\textbf{Search Cost Analysis for Offline/Online Deployment}
Since our automated optimization framework is a modeling-based search method, there is no need to profile the searched strategy each time, so our automatic optimization framework is low-cost.
In offline deployment, we can know all the multi-tenant deployment scenarios and can store the searched strategies in the device and use them directly when new requests appear.
For online deployment, \textit{GACER} could yield near-optimal schedule solutions within a short time, and our search cost is acceptable for tasks that care about throughput and are not sensitive to real-time.

\begin{figure*}[!t]
	\centering
	\captionsetup{justification=centering}
	\includegraphics[width=6.6in]{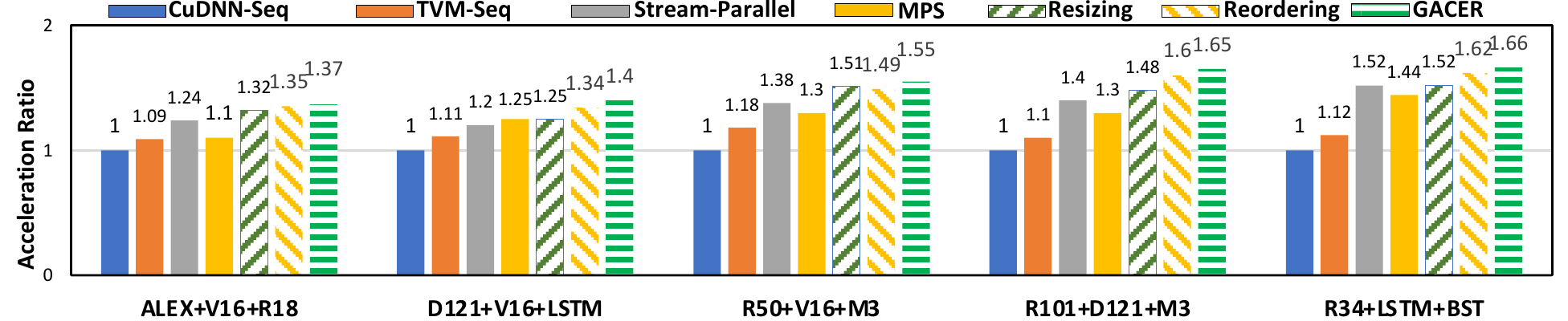}
	\caption{Runtime Performance of \textit{GACER} (with Titan V)}
	\vspace{-2mm}
	\label{fig:speedup}
\end{figure*}

\section{Performance Evaluation}
\label{sec:eval}

In this section, we evaluate the proposed system performance, including end-to-end speedup and GPU utilization enhancement targeting multi-tenant batched-job tasks \cite{kao2022magma}, in which each task has its own model batch size. 

\subsection{Experimental Setting and Metric}
\hspace{\parindent}
\textbf{Model Selection:}
We construct diverse multi-tenant DNN scenarios by leveraging three different types of applications collected from PyTorch: 
Vision models (including AlexNet (Alex), VGG16 (V16), ResNet18 (R18), ResNet34 (R34), ResNet50 (R50), ResNet101 (R101), MobileNetV3 (M3) and DenseNet (D121)), language model (LSTM \cite{yu2019review}), and transformer-based recommendation model (BST~\cite{chen2019BST}), which are commonly used for various applications such as photo auto-editing, image tagging, and video/voice processing.
These models have distinctive depths and varying numbers of operators, resulting in unique computational and memory requirements for each model.
Therefore, different model combinations based on the above models present varied resource utilization imbalances, thus requiring highly adaptive deployment and scheduling optimization strategies.


\vspace{1mm}
\textbf{Task and Benchmark:}
We consider the multi-tenant execution of the aforementioned DNNs and create workloads with varied batch sizes for each model.
For convolutional networks, we use an image scale of 224*224*3 with different batch size.
For the language model, we use the text dataset ML2020spring which is used for emotion classification.
And we use the Amazon dataset $Book$~\cite{zhou2018DIN} as the benchmark dataset for the recommendation model. 
For generality, we also evaluate four types of NVIDIA GPUs from desktop-level to high-end ones: 1080Ti, P6000, and Titan V. 

\vspace{1mm}
\textbf{Baseline Methods:}
Several popular baseline resource allocation optimization and scheduling strategies are considered.

$\bullet$ CuDNN-Seq \cite{CuDNN}: The default strategy of PyTorch + CuDNN, which runs the models sequentially.

$\bullet$ TVM-Seq \cite{chen2018tvm}: A operator-level optimization method to search for the optimal kernel for each operator.
    However, it can only run these kernels sequentially.

$\bullet$ Stream-Parallel \cite{Multi-Stream}: The concurrent execution strategy from native GPU multi-stream support. It assigns models to different streams and leverages the default GPU scheduler to schedule the execution sequence.

$\bullet$ MPS \cite{MPS}: We distribute the resources to each model based on the models' FLOPS. 


\vspace{1mm}
\textbf{Evaluation Metric:}
In our experiments, we use the end-to-end execution latency of multi-tenant DNNs and monitored GPU resource utilization as the optimization objectives.

\subsection{Speed-Up Evaluation}

We first compare the overall latency of the baselines and our methods \textit{GACER}. 
To demonstrate each design component's effectiveness, we decompose and evaluate our method step by step, i.e., using spatial granularity regulation (\textit{Spatial}), the method only using temporal granularity regulation (\textit{Temporal}), and the combined optimization (\textit{GACER}). 
The results are shown in Fig. \ref{fig:speedup}. 
All latency is normalized by the CuDNN-Seq baseline to show the relative acceleration ratio. 
We use five multi-tenant DNN combination settings, which cover a wide range of concurrency scenarios. 
For example, ALEX+VGG+R18 is a relatively simple one (10 $\sim$ 30 operators), and the number of operators of R101+D121+M3 can exceed 200.
R34+LSTM+BST is a complex multi-tenant scenario, which includes a vision model, language model, and recommendation model with diverse resource requirements.

\vspace{1mm}
\textbf{Overall \textit{GACER} Speed-up:}
Based on the results, it can be observed that our framework \textit{GACER} could consistently yield 1.37$\times$ $\sim$ 1.66$\times$ speed-up compared to the sequential baselines across all five model combinations. 
The Stream-Parallel solution also yields a certain speed-up than CuDNN-Seq, but the acceleration ratio is much less usually 1.24$\times$ $\sim$ 1.51$\times$.
Also, the MPS acceleration effect is very unstable, specifically due to that fixed resource allocation cannot satisfy many particularly unbalanced model workload scenarios.

\vspace{1mm}
\textbf{Speed-up with Spatial Granularity Regulation:}
The spatial granularity regulation is more advantageous for models with large operator workloads.
In particular, when the workloads of multiple parallel models are large, such as R50+V16+M3 in Fig. \ref{fig:speedup}, where both R50 and V16 both have large operator workloads, the decomposition of V16 can reduce more residue and obtain significant speedups.
On the contrary, for the combination of R34+LSTM+BST, the latter two models have a low SM occupation, so the workload decomposition has less optimization space for the residue and almost no performance gain compared to Stream-Parallel.

\vspace{1mm}
\textbf{Speed-up with Temporal Granularity Regulation:}
On the other hand, the temporal granularity regulation gives more significant speedup for model combinations with more layers.
The most obvious scenario is R101+D121+M3, in which all three models have more layers and complex operators in terms of timing and resource consumption.
Therefore, dividing them into multiple operator clusters allows for better residue reduction within each concurrent cluster.

\begin{figure}[!t]
	\centering
        \vspace{2mm}
	\includegraphics[width=3.3in]{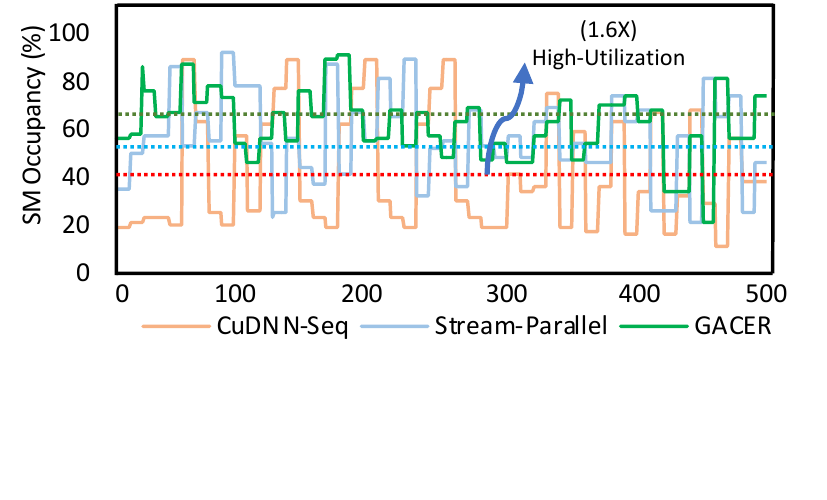}
	\vspace{-17mm}
	\caption{Analysis on GPU Utilization Enhancement}
	\vspace{-2mm}
	\label{fig:warps}
\end{figure}

\subsection{GPU Utilization Evaluation}

We further profiled and checked the GPU runtime statistics to analyze the overall GPU utilization within different scenarios.
We use the achieved SM Occupancy from NVIDIA NSight Profiler as an indicator metric of GPU utilization information.
Fig.~\ref{fig:warps} demonstrates the utilization statistics comparison between CuDNN-Seq, Stream-Parallel, and our method on the R101+D121+M3.
As is observed, our method obtains about 60\% utilization enhancement over the sequence method and almost 40\% enhancement than Stream-Parallel,  which is consistent with our speed-up performance.
\textit{GACER} runs with a more even utilization and has less inefficient intervals due to the fact that our method fills the residue well.

\begin{table*}[!t]
\setlength{\tabcolsep}{3mm}
	\centering
    \small
	\caption{GPU Generality Evaluation (ms)}\label{tab:generality}	
	\vspace{-3mm}
	\begin{tabular}{ccccccc}
		\hline
		Models        & C-P6000 & C-1080Ti & S-P6000 & S-1080Ti &\textit{GACER}-P6000 & \textit{GACER}-1080Ti \\
		\hline
		ALEX+V16+R18  & 18.74 & 19.56 & 14.99(\textbf{1.25x}) & 15.28(\textbf{1.28x}) & 13.48(\textbf{1.39x}) & 14.81(\textbf{1.32x}) \\
		\hline
		D121+V16+LSTM & 17.83 & 18.02 & 14.73(\textbf{1.21x}) & 15.27(\textbf{1.18x}) & 12.92(\textbf{1.38x}) & 13.54(\textbf{1.33x})\\
		\hline
		R50+V16+M3    & 28.54  & 32.88 & 20.88(\textbf{1.37x}) & 23.48(\textbf{1.40x}) & 19.02(\textbf{1.50x}) & 21.07(\textbf{1.56x})\\
		\hline
		R101+D121+M3  & 40.51 & 44.89 & 29.35(\textbf{1.38x}) & 32.06(\textbf{1.40x}) & 25.63(\textbf{1.58x}) & 27.37(\textbf{1.64x})\\
		\hline
	    R34+LSTM+BST & 12.35 & 14.50 & 8.23(\textbf{1.50x}) & 10.13(\textbf{1.43x}) & 7.97(\textbf{1.55x}) & 8.51(\textbf{1.70x})\\
	    \hline
	\end{tabular}
	\vspace{-3mm}
\end{table*}

\subsection{Framework Generality Analysis}
We then evaluate the generality of our method with different GPU platforms.
We perform operator profiling on different GPU platforms to obtain operator information lookup tables.
Then we use our search framework to get the optimal regulation solution.
We test five multi-tenant setups on different GPUs: the NVIDIA P6000 and the NVIDIA 1080Ti.
The P6000 GPU is the last version before Titan-V and has a slightly lower peak computing performance (12.6 vs. 14.9 TFLOPS). 
The 1080Ti is a relatively early platform, with a calculated peak performance of only 10.4 TFLPOS.
As the overall performance is shown in Table \ref{tab:generality}.
C is CuDNN-Seq, and S is Stream-Parallel.
Since these two platforms do not support MPS, it is not compared here. 
We set the batch size of each vision model to 8, the language model to 128, and the recommended model to 64, and we only test model inference here.
Our scheduling framework \textit{GACER} also yields significant performance gain (1.38$\times$ $\sim$ 1.58$\times$ acceleration on P6000 and 1.32$\times$ $\sim$ 1.70$\times$ acceleration on 1080Ti) on the different GPU platforms.



\subsection{Framework Granularity Awareness}

In this section, we conduct in-depth overhead profiling and analysis to demonstrate the full-spectrum system optimization trade-offs and how the \textit{GACER} framework decides coarse-to-fine temporal and spatial granularity to achieve optimal performance. 

\begin{figure}[!b]
	\centering
	\vspace{-4mm}
	\includegraphics[width=3.3in]{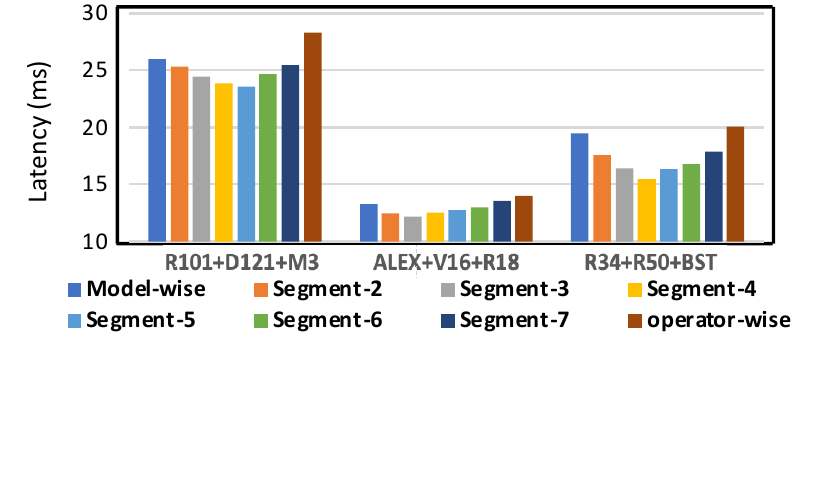}
	\vspace{-18mm}
	\caption{Different Temporal Granularity Performance}
	\vspace{-1mm}
	\label{fig:temporal_performance}
\end{figure}

\vspace{1mm}
\textbf{Temporal Granularity Trade-off:}
We compare the performance of multi-tenant DL computing under different scheduling granularity, including model-wise (Stream-Parallel), segment-wise (with a different number of pointers), and operator-wise. 
One interesting finding is that, as the scheduling granularity gets finer (from model to segment, and then operator), the latency performance tends to go through an improving and decreasing trend, forming a latency "sweet-zone" in the middle granularity as shown in Fig.~\ref{fig:temporal_performance}.

We choose three scenarios of model combinations and execute them in different scheduling granularity.
The "segment-2" means that we divide each model into 2 separate segments for scheduling.
As shown in Fig.~\ref{fig:temporal_performance}, although the "sweet-zone" always appears in the middle, the optimal granularity varies, which requires an adaptive granularity decision.
Based on the results, we can find that complex model combination tends to benefit more from more fine-grained segments to get the best performance, e.g. R101+D121+M3.
This is because they have more operators and thus the insertion of pointers usually has larger optimization space and freedom.
In contrast, the second combination is relatively simple and achieves optimal performance with the optimization of a small number of pointers. And too much scheduling can lead to performance degradation due to synchronization overhead.

\vspace{1mm}
\textbf{Spatial Granularity Trade-off:}
In this part, we analyze the parallel performance of some model combinations under different spatial granularity. 
For the baseline, we use two streams to parallel a VGG16 (V16) and ResNet18 (R18) model both with batch size 32.
In order to improve the granularity of resource allocation, we decompose the batch of some operators in two models separately.
We decompose all the convolution operators and the following Relu operators which are the main structure of these models
The performance of different cases is shown in the Table. \ref{tab:resource granularity}.
\begin{table}[!t]
    \setlength{\tabcolsep}{0.8mm}
    \centering
    \small
    \vspace{3mm}
    \caption{Different Spatial Granularity Performance}\label{tab:resource granularity}
    \vspace{-3mm}
    \begin{tabular}{cccccccc}
        \hline
         & S1 & S2 & S3 & S4 & S5 & Latency  \\
        \hline
        \circled{1} & V16(32) & R18(32)  & - & - & - & 80 ms \\
        \hline
        \circled{2} &V16(16) & V16(16)  & R18(32) & - & - & \textbf{66 ms} \\
        \hline
        \circled{3} &V16(24) & V16(8)  & R18(32) & - & - & 72 ms \\
        \hline
        \circled{4} &V16(32) & R18(16)  & R18(16) & - & - & 78 ms \\
        \hline
        \circled{5} &V16(8) & V16(8)  & V16(8) & V16(8) & R18(32) & 85 ms \\
        \hline
    \end{tabular}
    \captionsetup{font=footnotesize,skip=2pt}
    \caption*{Note: S1$\sim$ S5 are the different streams, the number in ``( )'' is the batch size, and each row is a combination of parallel models.}
    \vspace{-2mm}
\end{table}

The spatial granularity also has a "sweet zone" phenomenon, and the optimal strategy is not the most fine-grained.
This is because the decomposition and concatenation have the execution overhead and also increase more operators which could introduce more CPU operators issuing overhead. 
Another valuable phenomenon is the decomposition of the operator with a higher SM occupation can bring more benefits.
This phenomenon is particularly evident in the decomposition of the VGG convolutional layers. 
Most of the layers of V16 are computationally intensive convolutional layers, and these convolutional layers tend to occupy a large amount of SM resources during computation.
When computing the convolutional layers of V16, a large number of SMs are occupied and it is difficult for the operator from R18 to co-deploy, resulting in a large residue.
The operator decomposition allows the convolutional layers in V16 to be more flexible to parallel with the operators in R18, thus achieving a huge performance improvement, such as the case in Table. \ref{tab:resource granularity} \circled{2}.


\begin{table}[!t]
\setlength{\tabcolsep}{1.8mm}
	\centering
	\small
     \vspace{3mm}
	\caption{\textit{GACER} Search Overhead}\label{tab:scheduling_overhead}	
	\vspace{-3mm}
	\begin{tabular}{cccccc}
		\hline
		\#Search Rounds & 100 & 500 & 1000 & 2000 & 10000  \\
		\hline
		R34+V16+LSTM & 0.90s  & 4.61s & 9.65s & 20.56s & 1min43s\\
		\hline
		R50+V16+M3 &  1.60s & 7.9s & 16.1s & 33.0s & 2min52s \\
		\hline
		R34+LSTM+BST & 0.88s  & 4.8s & 9.8s & 20.3s & 1min30s \\
		\hline
	\end{tabular}
\end{table}

\subsection{Framework Overhead Analysis}
In this section, we analyze the overhead of our search algorithm, our framework could usually yield near-optimal schedule solutions within a short search time. 
The framework’s search running time overhead is demonstrated in Table \ref{tab:scheduling_overhead}. 
We profile the coordinate descent search with different search rounds from 100 to 10000.
The results show that the running overhead of our framework stays in the range of several seconds to at most a few minutes.
Also, since it is possible to pre-execute this automated search algorithm offline for a given multi-tenant scenario, we consider this offline tuning overhead to be very acceptable.
Such a search overhead is also acceptable for some online tasks that are not very real-time, such as training.

\section{Conclusion}
\label{sec:conc}
In this work, we focus on multi-tenant deep learning for GPU-based systems and  reveal several computing issues.
We find that the granularity of computing management in both the spatial and temporal management domains is extremely important.
However, there is a lack of a library and framework for granularity optimization. 
Based on the optimization of granularity, we proposed \textit{GACER}, an automated optimization framework that provides high-utilization, high-throughput, and low-latency multi-tenant deep learning computing support.
This framework is a revolutionary approach to multi-tenant deployment and can provide a solid foundation for the future development of multi-tenant deep learning.
\setcounter{secnumdepth}{0}
\bibliographystyle{unsrtnat}
\vspace{-2mm}
\bibliography{reference.bib}

\begin{thebibliography}{31}
\providecommand{\natexlab}[1]{#1}
\providecommand{\url}[1]{\texttt{#1}}
\expandafter\ifx\csname urlstyle\endcsname\relax
  \providecommand{\doi}[1]{doi: #1}\else
  \providecommand{\doi}{doi: \begingroup \urlstyle{rm}\Url}\fi

\bibitem[Owens et~al.(2008)Owens, Houston, Luebke, Green, Stone, and
  Phillips]{owens2008gpu}
John~D Owens, Mike Houston, David Luebke, Simon Green, John~E Stone, and
  James~C Phillips.
\newblock Gpu computing.
\newblock \emph{Proceedings of the IEEE}, 96\penalty0 (5):\penalty0 879--899,
  2008.

\bibitem[Keckler et~al.(2011)Keckler, Dally, Khailany, Garland, and
  Glasco]{keckler2011gpus}
Stephen~W Keckler, William~J Dally, Brucek Khailany, Michael Garland, and David
  Glasco.
\newblock Gpus and the future of parallel computing.
\newblock \emph{IEEE micro}, 31\penalty0 (5):\penalty0 7--17, 2011.

\bibitem[Reports(2021)]{gpu_report}
Market Reports.
\newblock Global data center accelerator market size, status and forecast
  2020-2025, 2021.
\newblock
  \url{https://www.mynewsdesk.com/brandessence/pressreleases/data-center-accelerator-market-size-2021-cagr-38-dot-7-percent-3112488}.

\bibitem[Jia et~al.(2019)Jia, Thomas, Warszawski, Gao, Zaharia, and
  Aiken]{jia2019optimizing}
Zhihao Jia, James Thomas, Todd Warszawski, Mingyu Gao, Matei Zaharia, and Alex
  Aiken.
\newblock Optimizing dnn computation with relaxed graph substitutions.
\newblock \emph{Proceedings of Machine Learning and Systems}, 1:\penalty0
  27--39, 2019.

\bibitem[Ding et~al.(2021)Ding, Zhu, Jia, Pekhimenko, and Han]{ding2021ios}
Yaoyao Ding, Ligeng Zhu, Zhihao Jia, Gennady Pekhimenko, and Song Han.
\newblock Ios: Inter-operator scheduler for cnn acceleration.
\newblock \emph{Proceedings of Machine Learning and Systems}, 3:\penalty0
  167--180, 2021.

\bibitem[Lin et~al.(2019)Lin, Ji, Yan, Zhang, Cao, Ye, Huang, and
  Doermann]{lin2019towards}
Shaohui Lin, Rongrong Ji, Chenqian Yan, Baochang Zhang, Liujuan Cao, Qixiang
  Ye, Feiyue Huang, and David Doermann.
\newblock Towards optimal structured cnn pruning via generative adversarial
  learning.
\newblock In \emph{Proceedings of the IEEE/CVF Conference on Computer Vision
  and Pattern Recognition}, pages 2790--2799, 2019.

\bibitem[Howard et~al.(2017)Howard, Zhu, Chen, Kalenichenko, Wang, Weyand,
  Andreetto, and Adam]{howard2017mobilenets}
Andrew~G Howard, Menglong Zhu, Bo~Chen, Dmitry Kalenichenko, Weijun Wang,
  Tobias Weyand, Marco Andreetto, and Hartwig Adam.
\newblock Mobilenets: Efficient convolutional neural networks for mobile vision
  applications.
\newblock \emph{arXiv preprint arXiv:1704.04861}, 2017.

\bibitem[Liu et~al.(2019)Liu, Liu, Tang, Yu, Wang, and Shi]{liu2019edge}
Shaoshan Liu, Liangkai Liu, Jie Tang, Bo~Yu, Yifan Wang, and Weisong Shi.
\newblock Edge computing for autonomous driving: Opportunities and challenges.
\newblock \emph{Proceedings of the IEEE}, 107\penalty0 (8):\penalty0
  1697--1716, 2019.

\bibitem[Mystakidis(2022)]{mystakidis2022metaverse}
Stylianos Mystakidis.
\newblock Metaverse.
\newblock \emph{Encyclopedia}, 2\penalty0 (1):\penalty0 486--497, 2022.

\bibitem[NVIDIA(2021)]{MPS}
NVIDIA.
\newblock {Multi-Process Service}, 2021.
\newblock URL
  \url{https://docs.nvidia.com/deploy/pdf/CUDA_Multi_Process_Service_Overview.pdf}.

\bibitem[NVIDIA(2020{\natexlab{a}})]{Multi-Stream}
NVIDIA.
\newblock {Multi-Stream}, 2020{\natexlab{a}}.
\newblock URL
  \url{https://on-demand.gputechconf.com/gtc/2014/presentations/S4158-cuda-streams-best-practices-common-pitfalls.pdf}.

\bibitem[NVIDIA(2020{\natexlab{b}})]{MIG}
NVIDIA.
\newblock Nvidia multi instance gpu (mig), 2020{\natexlab{b}}.
\newblock URL \url{https: //docs.nvidia.com/datacenter/tesla/mig-user-guide/}.

\bibitem[Yu et~al.(2022)Yu, Yu, Xu, Wang, Zhang, Li, Bray, Liu, and
  Chen]{yu2022powering}
Yongbo Yu, Fuxun Yu, Zirui Xu, Di~Wang, Minjia Zhang, Ang Li, Shawn Bray,
  Chenchen Liu, and Xiang Chen.
\newblock Powering multi-task federated learning with competitive gpu resource
  sharing.
\newblock In \emph{Companion Proceedings of the Web Conference 2022}, pages
  567--571, 2022.

\bibitem[Choi et~al.(2021{\natexlab{a}})Choi, Kim, and Rhu]{choi2021lazy}
Yujeong Choi, Yunseong Kim, and Minsoo Rhu.
\newblock Lazy batching: An sla-aware batching system for cloud machine
  learning inference.
\newblock In \emph{2021 IEEE International Symposium on High-Performance
  Computer Architecture (HPCA)}, pages 493--506. IEEE, 2021{\natexlab{a}}.

\bibitem[Koller et~al.(2019)Koller, Camgoz, Ney, and Bowden]{koller2019weakly}
Oscar Koller, Necati~Cihan Camgoz, Hermann Ney, and Richard Bowden.
\newblock Weakly supervised learning with multi-stream cnn-lstm-hmms to
  discover sequential parallelism in sign language videos.
\newblock \emph{IEEE transactions on pattern analysis and machine
  intelligence}, 42\penalty0 (9):\penalty0 2306--2320, 2019.

\bibitem[Kwon et~al.(2020)Kwon, Yu, Jeong, and Chun]{kwon2020nimble}
Woosuk Kwon, Gyeong-In Yu, Eunji Jeong, and Byung-Gon Chun.
\newblock Nimble: Lightweight and parallel gpu task scheduling for deep
  learning.
\newblock \emph{Advances in Neural Information Processing Systems},
  33:\penalty0 8343--8354, 2020.

\bibitem[Dhakal et~al.(2020)Dhakal, Kulkarni, and
  Ramakrishnan]{dhakal2020gslice}
Aditya Dhakal, Sameer~G Kulkarni, and KK~Ramakrishnan.
\newblock Gslice: Controlled spatial sharing of gpus for a scalable inference
  platform.
\newblock In \emph{Proceedings of the 11th ACM Symposium on Cloud Computing},
  pages 492--506, 2020.

\bibitem[Yu et~al.(2021)Yu, Bray, Wang, Shangguan, Tang, Liu, and
  Chen]{yu2021automated}
Fuxun Yu, Shawn Bray, Di~Wang, Longfei Shangguan, Xulong Tang, Chenchen Liu,
  and Xiang Chen.
\newblock Automated runtime-aware scheduling for multi-tenant dnn inference on
  gpu.
\newblock In \emph{2021 IEEE/ACM International Conference On Computer Aided
  Design (ICCAD)}, pages 1--9. IEEE, 2021.

\bibitem[Xu et~al.(2016)Xu, Jeon, Kim, Ro, and Annavaram]{xu2016warped}
Qiumin Xu, Hyeran Jeon, Keunsoo Kim, Won~Woo Ro, and Murali Annavaram.
\newblock Warped-slicer: Efficient intra-sm slicing through dynamic resource
  partitioning for gpu multiprogramming.
\newblock In \emph{2016 ACM/IEEE 43rd Annual International Symposium on
  Computer Architecture (ISCA)}, pages 230--242. IEEE, 2016.

\bibitem[Choi et~al.(2021{\natexlab{b}})Choi, Lee, Kim, Park, Kwon, and
  Huh]{choi2021multi}
Seungbeom Choi, Sunho Lee, Yeonjae Kim, Jongse Park, Youngjin Kwon, and Jaehyuk
  Huh.
\newblock Multi-model machine learning inference serving with gpu spatial
  partitioning.
\newblock \emph{arXiv preprint arXiv:2109.01611}, 2021{\natexlab{b}}.

\bibitem[Yu and Chowdhury(2020)]{yu2020salus}
Peifeng Yu and Mosharaf Chowdhury.
\newblock Fine-grained gpu sharing primitives for deep learning applications.
\newblock \emph{Proceedings of Machine Learning and Systems}, 2:\penalty0
  98--111, 2020.

\bibitem[Zhang et~al.(2019)Zhang, Vance, Zhang, Rashid, and
  Wang]{zhang2019edgebatch}
Daniel Zhang, Nathan Vance, Yang Zhang, Md~Tahmid Rashid, and Dong Wang.
\newblock Edgebatch: Towards ai-empowered optimal task batching in intelligent
  edge systems.
\newblock In \emph{2019 IEEE Real-Time Systems Symposium (RTSS)}, pages
  366--379. IEEE, 2019.

\bibitem[Huang et~al.(2019)Huang, Cheng, Bapna, Firat, Chen, Chen, Lee, Ngiam,
  Le, Wu, et~al.]{huang2019gpipe}
Yanping Huang, Youlong Cheng, Ankur Bapna, Orhan Firat, Dehao Chen, Mia Chen,
  HyoukJoong Lee, Jiquan Ngiam, Quoc~V Le, Yonghui Wu, et~al.
\newblock Gpipe: Efficient training of giant neural networks using pipeline
  parallelism.
\newblock \emph{Advances in neural information processing systems}, 32, 2019.

\bibitem[Chen et~al.(2018)Chen, Moreau, Jiang, Zheng, Yan, Shen, Cowan, Wang,
  Hu, Ceze, et~al.]{chen2018tvm}
Tianqi Chen, Thierry Moreau, Ziheng Jiang, Lianmin Zheng, Eddie Yan, Haichen
  Shen, Meghan Cowan, Leyuan Wang, Yuwei Hu, Luis Ceze, et~al.
\newblock $\{$TVM$\}$: An automated $\{$End-to-End$\}$ optimizing compiler for
  deep learning.
\newblock In \emph{13th USENIX Symposium on Operating Systems Design and
  Implementation (OSDI 18)}, pages 578--594, 2018.

\bibitem[NVIDIA(2020{\natexlab{c}})]{Nsight}
NVIDIA.
\newblock {NVIDIA Nsight Systems}, 2020{\natexlab{c}}.
\newblock URL
  \url{https://on-demand.gputechconf.com/gtc/2014/presentations/S4158-cuda-streams-best-practices-common-pitfalls.pdf}.

\bibitem[META(2020)]{pytorch}
META.
\newblock Pytorch, 2020.
\newblock URL \url{https://pytorch.org}.

\bibitem[Kao and Krishna(2022)]{kao2022magma}
Sheng-Chun Kao and Tushar Krishna.
\newblock Magma: An optimization framework for mapping multiple dnns on
  multiple accelerator cores.
\newblock In \emph{2022 IEEE International Symposium on High-Performance
  Computer Architecture (HPCA)}, pages 814--830. IEEE, 2022.

\bibitem[Yu et~al.(2019)Yu, Si, Hu, and Zhang]{yu2019review}
Yong Yu, Xiaosheng Si, Changhua Hu, and Jianxun Zhang.
\newblock A review of recurrent neural networks: Lstm cells and network
  architectures.
\newblock \emph{Neural Computation}, 31\penalty0 (7):\penalty0 1235--1270,
  2019.

\bibitem[Chen et~al.(2019)Chen, Zhao, Li, Huang, and Ou]{chen2019BST}
Qiwei Chen, Huan Zhao, Wei Li, Pipei Huang, and Wenwu Ou.
\newblock Behavior sequence transformer for e-commerce recommendation in
  alibaba.
\newblock In \emph{Proceedings of the 1st International Workshop on Deep
  Learning Practice for High-Dimensional Sparse Data}, pages 1--4, 2019.

\bibitem[Zhou et~al.(2018)Zhou, Zhu, Song, Fan, Zhu, Ma, Yan, Jin, Li, and
  Gai]{zhou2018DIN}
Guorui Zhou, Xiaoqiang Zhu, Chenru Song, Ying Fan, Han Zhu, Xiao Ma, Yanghui
  Yan, Junqi Jin, Han Li, and Kun Gai.
\newblock Deep interest network for click-through rate prediction.
\newblock In \emph{Proceedings of the 24th ACM SIGKDD international conference
  on knowledge discovery \& data mining}, pages 1059--1068, 2018.

\bibitem[NVIDIA(2020{\natexlab{d}})]{CuDNN}
NVIDIA.
\newblock {Nvidia CuDNN Documentation,}, 2020{\natexlab{d}}.
\newblock URL
  \url{https://docs.nvidia.com/deeplearning/cudnn/developer-guide/index.html}.

\end{thebibliography}
\end{document}